\documentclass[usenatbib,useAMS]{mn2e}
\newcommand{\si}[1]{\ensuremath{_{\textrm{\tiny{#1}}}}}

\usepackage{graphicx}

\title[]{Multiwavelength View of the Inner Spiral of NGC\,1365}
\author[]{Iskra V. Strateva$^{1}$\thanks{E-mail: strateva@mpe.mpg.de} and S. Komossa$^{1}$\footnotemark[1]\\
$^{1}$Max Planck Institut f\"ur extraterrestrische Physik, Giessenbachstrasse, 85748 Garching, Germany\\
}
\begin{document}

\date{Received 2009 February 25}

\pagerange{\pageref{firstpage}--\pageref{lastpage}} \pubyear{2002}

\maketitle

\label{firstpage}

\begin{abstract}
We study the extended nuclear emission of the starburst galaxy NGC\,1365. A weak obscured AGN and a strong starburst both contribute to the observed X-ray, optical, infrared, and radio emission in the inner 2\,kpc. The X-ray emission is spatially resolved, allowing comparison with multiwavelength data that highlights the structures dominating the nuclear region: the AGN, the nuclear spiral, the circumnuclear starburst ring, and nuclear outflow. The ultrasoft X-ray emission $\la0.5$\,keV is spatially coincident with the conical outflow traced by higher excitation optical emission lines like [O III] and [Ne III]. The strong starburst concentrated in super-star clusters in a circumnuclear ring with radius $\approx$1\,kpc dominates the 0.5--1.5\,keV emission and is visible in radio, molecular CO, and infrared maps of the central kiloparsec. The hard (2--10\,keV) emission is dominated by the obscured AGN, but also contributes to the emission from relatively old ($\sim7$\,Myr) but still enshrouded in dust and extremely massive ($10^7$\,M$_{\odot}$) super-star clusters \citep{G08}, hidden from view in the optical and soft X-ray bands. In the Appendix we present the X-ray spectroscopy and photometry of BL Lac MS\,0331.3$-$3629, a high-energy peaked BL Lac candidate at $z=0.308$, serendipitously detected in one \emph{Chandra} and five \emph{XMM-Newton} observations of NGC\,1365.
\end{abstract}


\section{Introduction}

\begin{figure*}
\includegraphics[width=17cm]{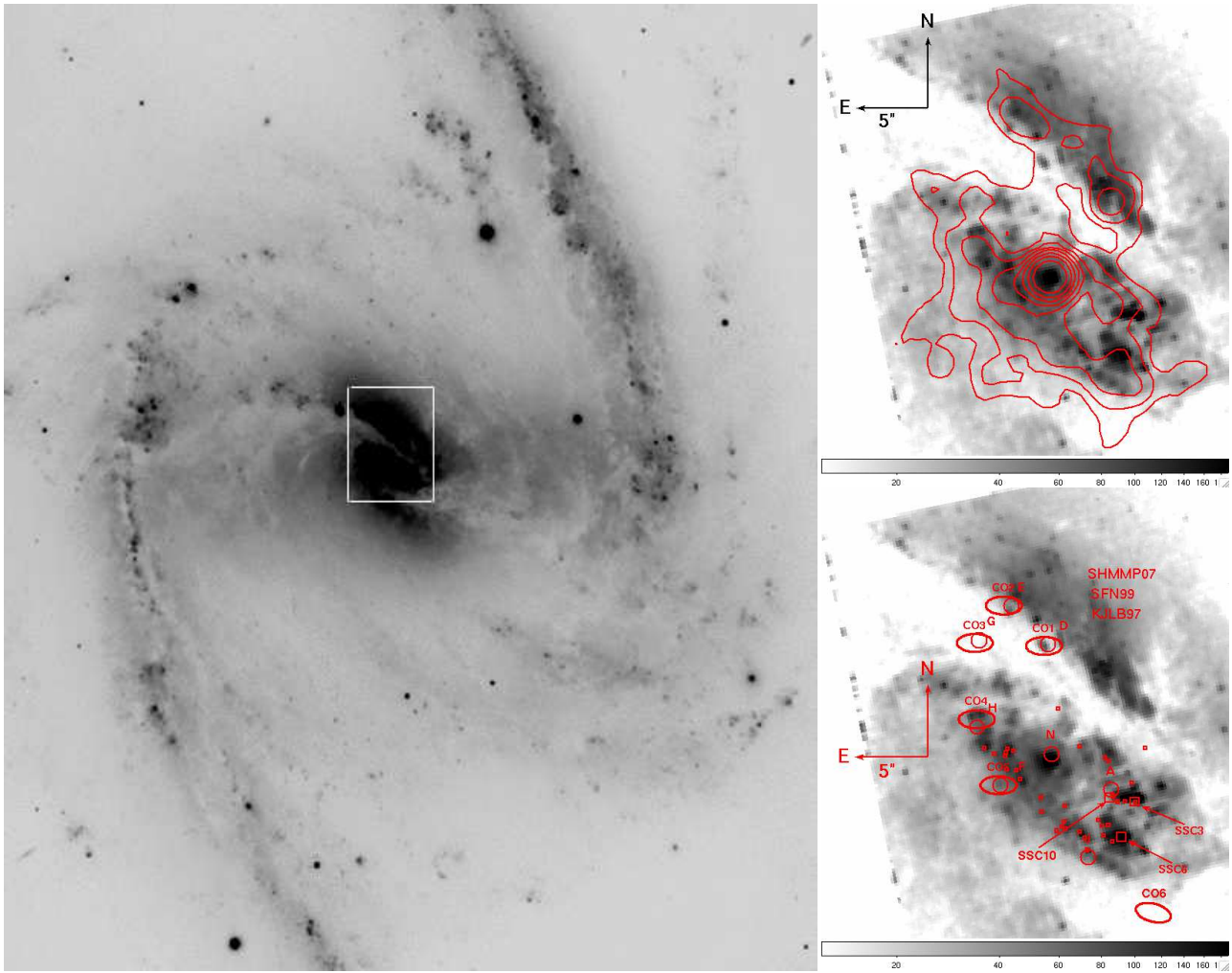}
\caption{\emph{Left:} Optical VLT image of NGC\,1365 with the inner spiral region marked. \emph{Top right:} \emph{HST} WFPC2 F555W image of the inner spiral of NGC\,1365 with the \hbox{0.3--10\,keV} band  logarithmically-spaced contours from the 2002 \emph{Chandra} observation overlaid in red. \emph{Bottom right:} The circumnuclear starburst ring of NGC\,1365 is shown here over the same \emph{HST} image, traced by the optical super-star clusters \citep[the SSCs are denoted by squares;][KJL97]{K97}, the radio hotsplots \citep[circles;][SFN99]{St99}, and the $^{12}$CO molecular hotspots \citep[ellipses;][SHMMP07]{S07}.
\label{HSTChandra}}
\end{figure*}

\emph{Chandra}'s excellent imaging resolution allows us an unprecedented glimpse of the spatially resolved X-ray emission of nearby galaxies. Unlike the simple nuclear point-source emission characterizing the X-ray view of luminous quasars, nearby starburst galaxies and active galactic nuclei (AGNs) show a plethora of diverse X-ray features: off-center point source emission from X-ray binaries, emission from supernova remnants, diffuse collisionally ionized emission from galactic halos, emission from photoionized narrow line regions (NLRs), as well as the X-ray emission of jets and outflows, including starburst-driven superwinds \citep[e.g.,][]{n4631,fabbiano02,wang03,Ogle03,Strickland04,Read06}. When combined with optical, radio, and infrared emission maps, the spatially resolved X-ray emission allows us to identify important processes associated with the birth and deaths of stars, the size, place, and composition of regions of recent star formation, as well as to gauge the influence of the AGN on the nuclear regions of galaxies.

The barred supergiant\footnote{The $B=25^m$\,arcsec$^{-2}$ isophotal radius of NGC\,1365 is 30\,kpc.} spiral NGC\,1365 has been the target of extensive studies from the radio \citep[e.g.,][and references therein]{Sand95,St99} to the X-ray band \citep[e.g.,][]{Turner93,Iyomoto97,KS98,R05,R05b,R07}. The multiwavelength data on NGC\,1365 was recently reviewed in \citet{L99}. In the X-ray band, there are at least three distinct emission components. The obscured active nucleus \citep{Iyomoto97,KS98,R05,R05b,R07}, the soft, extended emission \citep{Turner93,KS98,Wang09}, and the X-ray point source populations \citep{Turner93,KS98,Soria07,SK09}. 

NGC 1365 is special in the sense that several of its multi-wavelength properties are, at first glance, unusual.
(1) The degree of excitation of {\em narrow} emission-lines detected in optical spectra is actually lower in the core region of NGC\,1365 than in the extranuclear regions \citep[e.g.,][]{Edmunds88,Schulz99,Veilleux}. 
(2) While this could be explained by strong core extinction, a {\em broad} component nevertheless shows up in the Balmer lines \citep{Veron,EP82,Schulz99}. The width of the broad-line component is $\sim$1600--1900\,km\,s$^{-1}$ and would formally imply a classification as narrow-line Seyfert 1 galaxy, even though the other properties of NGC\,1365 do not match that classification.
(3) Compared to other samples of Seyfert galaxies \citep{Ward88}, the optical and IR properties of NGC\,1365 appear to be off the correlations and trends in multi-wavelength diagrams \citep{KS98}.
(4) To this adds the recent finding that the X-ray obscuration of NGC\,1365, which is unusually variable on very short timescales \citep[e.g.,][]{R05}, is likely caused by the broad-line region clouds \citep{R08}.

With the key goals in mind of spatially resolving the soft X-ray emission seen with ROSAT and disentangling AGN and starburst components, getting new clues on the multi-wavelength properties of NGC 1365, and studying its population of X-ray sources including the luminous source NGC\,1365-X1, we obtained a Chandra ACIS-S observation of NGC\,1365 on MPE guaranteed time.

In this paper we report on the extended X-ray emission of NGC\,1365. We show that the 0.5--10\,keV extended nuclear emission originates in the star-forming ring and nuclear spiral in the inner 2\,kpc of NGC\,1365, as traced by radio hotspots, molecular CO emission, mid-infrared (MIR) emission, and optically detected super star clusters, while ultrasoft ($<$0.5\,keV) emission traces a nuclear outflow cone \citep{SBB91,HL96,L99,Veilleux}. Throughout this paper we assume a luminosity distance\footnote{NASA/IPAC Extragalactic Database, http://nedwww.ipac.caltech.edu/.} of 21\,Mpc to NGC\,1365. All quoted flux and luminosity uncertainties are 90\% confidence limits, unless noted otherwise. 

\begin{figure*}
\includegraphics[width=17cm]{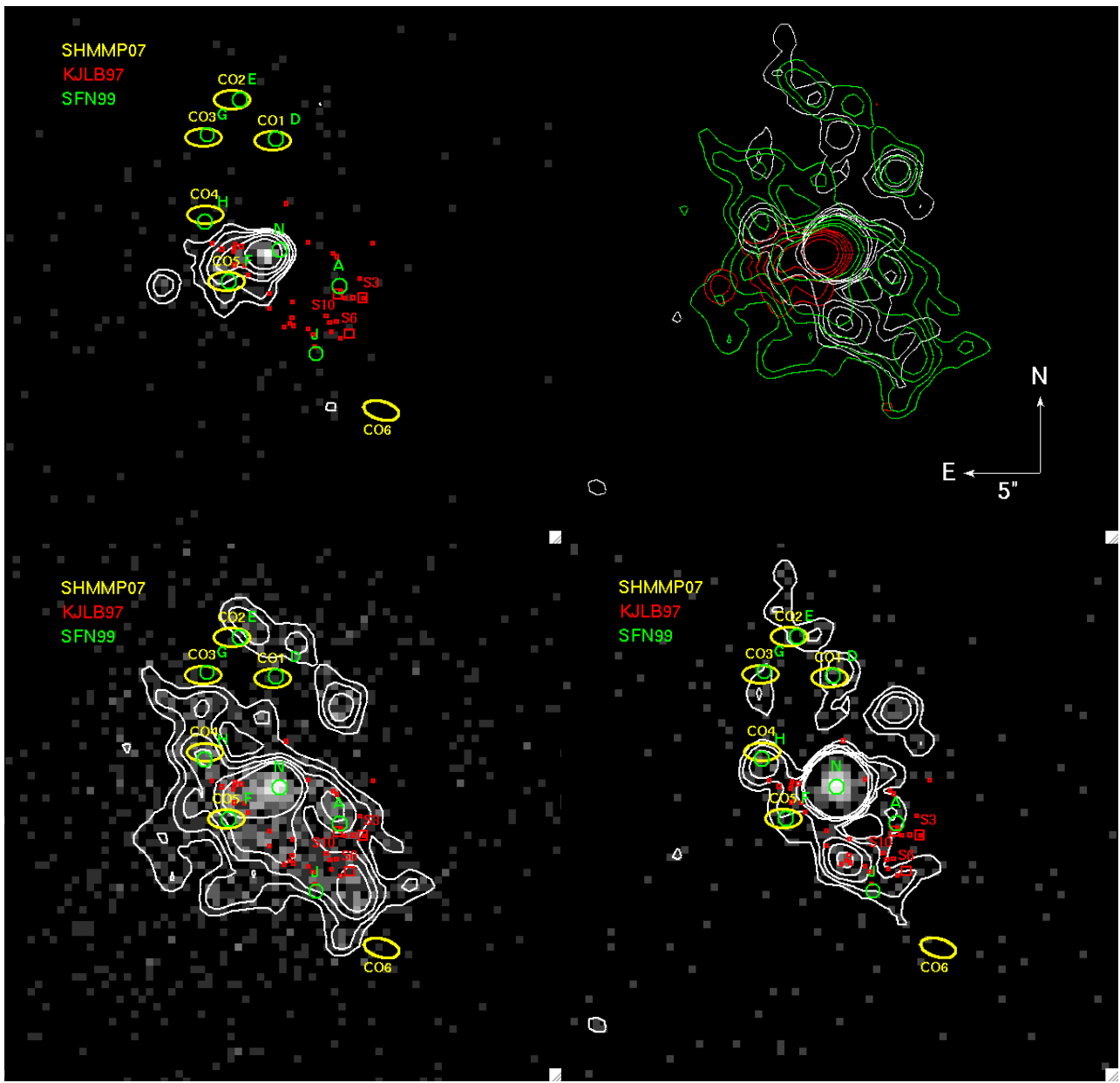}
\caption{Logarithmically spaced X-ray contours and images during the 2002 \emph{Chandra} observation in the ultrasoft (\hbox{0.2--0.5\,keV}; {\it top left}), the soft (\hbox{0.5--1.5\,keV}; {\it bottom left}), and hard band (\hbox{1.5--10\,keV}; {\it bottom right}). The highest level contours are selected to exclude the AGN point sources in each band, while lowest level contours are selected to exclude background emission. The top right panel overplots the soft (red), medium (green), and hard band (white) emission contours. The  circumnuclear star-forming ring of NGC\,1365 is also shown here traced by the optical super-star clusters \citep[the SSCs, denoted by red squares;][KJL97]{K97}, the radio hotsplots \citep[green circles;][SFN99]{St99}, and the $^{12}$CO molecular hotspots \citep[yellow ellipses;][SHMMP07]{S07}. The soft band contours are extended in the direction of the radio jet (radio hotspot F), and the  [O III] $\lambda$5007 outflow cone of \citet{HL96} and \citet{L99}. 
\label{obs3554}}
\end{figure*} 

\section{The Extended X-ray Emission of NGC\,1365}

\begin{table}
\begin{minipage}{8cm}
\caption{NGC\,1365: X-ray Observation Summary. T$\si{eff}$ is the effective exposure time after corrections for flaring and chip position.}
\begin{tabular}{@{}ccc@{}}
\hline
ObsID & MJD & T$\si{eff}$ [10$^3$\,s]\\
\hline
\emph{Chandra} 3554	& 52632 &  9.9\,(ACIS-S3) \\
\emph{Chandra} 6871 	& 53835 &  13.4\,(ACIS-S3) \\
\emph{Chandra} 6872	& 53837 &  14.6\,(ACIS-S3) \\
\emph{Chandra} 6873	& 53840 &  14.6\,(ACIS-S3) \\
\emph{Chandra} 6868	& 53843 & 14.6\,(ACIS-S3)\\
\emph{Chandra} 6869	& 53845 & 14.6\,(ACIS-S3)\\
\emph{Chandra} 6870	& 53848  & 15.5\,(ACIS-S3)\\
XMM 0151370101 		& 52655 &14.4\,(EPIC pn) 17.6\,(MOS) \\ 
XMM 0151370201 		& 52679 &  2.5\,(EPIC pn) 5.4\,(MOS) \\ 
XMM 0151370701 		& 52864 &   6.6\,(EPIC pn) 8.2\,(MOS)\\ 
XMM 0205590301 		& 53021 & 49.0\,(EPIC pn) 56.5(MOS)\\ 
XMM 0205590401		& 53210 &  30.0\,(EPIC pn) 44.6\,(MOS)\\
\hline
\end{tabular}
\end{minipage}
\label{obsTab}
\end{table}

\subsection{Observation Summary}
\label{obssummary}

NGC\,1365 was first imaged by \emph{Chandra} in 2002, December. The observation shows diffuse extended nuclear emission in the inner $20''\times20''$ ($\approx2\times2$\,kpc) and 23 off-nuclear point sources in the central $\sim6'\times6'$ region \citep[][hereafter SK09]{SK09}. Between 2004 and 2006, NGC\,1365 was the target of extensive X-ray campaigns which resulted in six additional \emph{Chandra} and five \emph{XMM-Newton} observations (with EPIC-pn exposures of 5.4\,ks to 56.5\,ks); for more details of the exposure times and dates of these observations, see Table~\ref{obsTab}.  We processed all seven \emph{Chandra} observations with CIAO v.3.4. The 14.6\,ks Obs.\,ID\,3554 observation is reduced to 9.9\,ks after the exclusion of a 4.7\,ks flare. The remaining 6 \emph{Chandra} observations are free of flaring, with effective times ranging between 13.4 and 15.5\,ks. 

Using both \emph{Chandra} and \emph{XMM-Newton} we can study in detail the extended emission of NGC\,1365, through both spectroscopy and spatially resolved imaging, excluding the absorbed and highly-variable nuclear point source described in detail by e.g., \citet{R05}. In this paper we present visual results based on our original 2002 \emph{Chandra} exposure, while using the full \emph{Chandra} and \emph{XMM-Newton} datasets in the analysis.

\begin{figure}
\includegraphics[width=8cm]{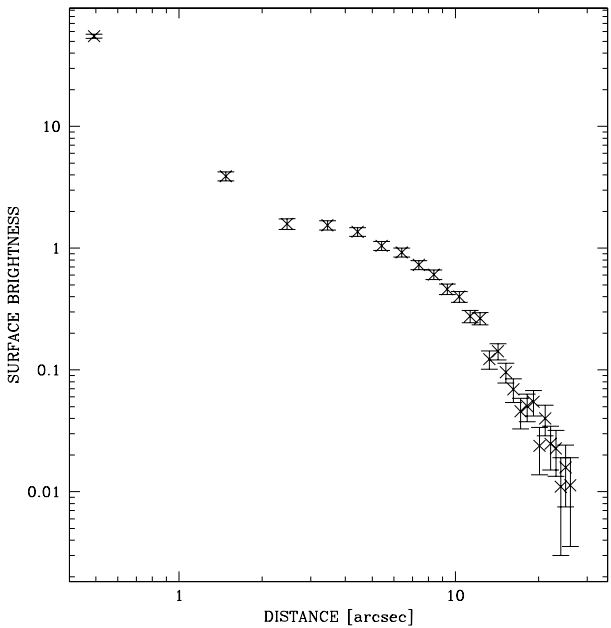}
\caption{Radial profile of the X-ray emission of NGC\,1365 from 2002. The AGN point source is clearly visible in the inner $\sim1-2''$; the X-ray emission associated with the inner spiral extends to at least $20''$. 
\label{rprof}}
\end{figure}

\begin{figure}
\includegraphics[angle=270,width=8cm]{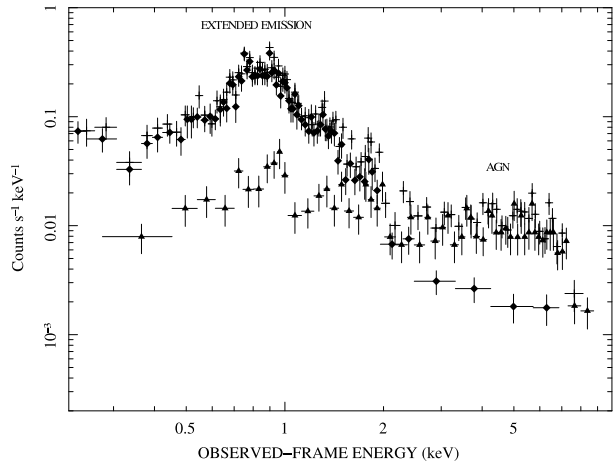}
\caption{\emph{Chandra} spectra of the AGN point source ($<2''$; rombs), the extended emission ($2''<r<40''$, see text; triangles) and the total emission (crosses). The AGN emission dominates above $\sim2.5$\,keV, the extended emission dominates below $\sim1.5$\,keV.
  \label{AllSpec}}
\end{figure}

\subsection{The Inner Spiral and Outflow Cone of NGC\,1365}

Kinematically, the nuclear region of NGC\,1365 shows two distinct parts: (1) the rotation field of the inner spiral, and (2) the outflowing cone component. The inner spiral is well traced by the optical/infrared (IR) continuum, as well as low excitation lines like the Balmer lines, and the forbidden [N II], [S II], and [O II] lines. The inner spiral has two prominent dust lanes (clearly visible on the HST image in Fig.~\ref{HSTChandra}), which can be traced by dense molecular gas, e.g., amonia \citep[NH$_3$;][]{Ott05}.  The outflowing cone component is traced by higher excitation lines like [O III]$\,\lambda\lambda$5007,4958, [Ne III]$\,\lambda$3868, and He~II$\,\lambda$4686 \citep{HL96,L99} and can be seen on Figs.~19 and 22 of \citet{L99}. \citet{Veilleux} constructed a spatially resolved map of the narrow-line-region flux ratios ([O III]/H$\beta$ and  [N II]/H$\alpha$) of the central few kiloparsecs of NGC\,1365. They concluded that starburst-like photoionization dominates the nuclear spiral, with the exception of the region associated with the strongest [O III] emission, which, together with the region outside 2\,kpc, has the characteristic ratios of AGN photoionization.

\subsection{Spatially-Resolved X-ray Emission}
\label{spatial}

The left panel of Figure~\ref{HSTChandra} gives a panoramic view of the NGC\,1365 spiral, where we have outlined the nuclear region that we study in detail. The top right panel of Figure~\ref{HSTChandra} shows the \hbox{0.3--10\,keV} X-ray emission contours from the 2002 \emph{Chandra} exposure overlaid on the F555W WFPC2 image of the inner spiral\footnote{For the multiwavelength comparison we have applied a $\sim1''$ correction to the X-ray positions requiring that the ($>$0.5\,keV)  X-ray emission peak coincides with the position of the optical emission peak and the radio center N.}. The \hbox{0.3--10\,keV} X-ray emission traces well the nuclear spiral component, suggesting that the extended X-ray emission is associated with the circumnuclear starburst ring with projected size $\sim20''\times10''$ ($\sim$2\,kpc$\times$1\,kpc). The nuclear starburst ring is traced completely by radio 3\,cm and 6\,cm emission \citep{FN98,St99}, and partially by molecular CO emission \citep{S07} and optically detected super-star clusters \citep[SSCs;][]{K97}, as shown in the bottom right panel of Figure~\ref{HSTChandra}.

The spatial distribution of the hard ($>$1.5\,keV) and ultrasoft ($<$0.5\,keV) \emph{Chandra} X-ray photons provide further clues to the origin of the extended X-ray emission. The $2''$-smoothed X-ray images and contours of the extended emission from the 2002 exposure are presented in Figure~\ref{obs3554} in three different energy bands. The \hbox{0.2--0.5\,keV}  X-ray emission shown in the upper left panel is not spatially associated with the circumnuclear ring; instead it traces the outflow cone seen in higher optical excitation lines like [O III]. The ultrasoft X-ray emission also coincides with the direction of the probable nuclear jet \citep[extending from the nucleus, N, to radio hotspot F;][]{St99}, peaking $0.8''$ SW from the centers of the soft- and hard-band emission (which coincide with the radio center N by construction; top right panel of Figure~\ref{obs3554}). Following the spatial coincidence between the  radio, [O III], and ultrasoft X-ray emission, which is confirmed by the emission maps of the remaining 6 \emph{Chandra} observations, we suggest that the ultrasoft emission is either part of an X-ray jet or, more likely, originates in the wide-angle outflow. 

\begin{figure}
\includegraphics[angle=270,width=9cm]{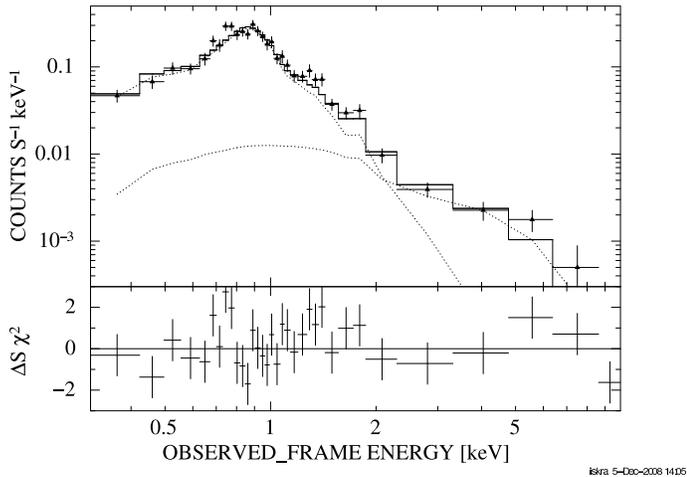}
\caption{Best fit 0.3-10\,keV spectrum for \emph{Chandra} Obs.\,ID~3554. The model fit includes a thermal plasma (\emph{apec} in XSPEC) with a temperature of $\sim$0.7\,keV and sub-solar abundance, and a power-law tail with a photon index $\Gamma=1$. The two model components are shown with dotted lines in the top panel, the fit residuals are shown in a separate panel below.
\label{ChandraFullSpec}}
\end{figure}

The hard (\hbox{1.5--10\,keV}) band emission shown in the lower right panel of Figure~\ref{obs3554} coincides with some of the radio and molecular hotspots and SSCs. The converse is not true -- not all the SSCs have hard X-ray counterparts  -- the brightest SSCs (S10 and S3), for example, do not correspond to hard X-ray peaks. The general impression on this and the remaining 6 \emph{Chandra} exposures of the same region is that the hard X-rays trace well the circumnuclear starburst ring vissible in the optical, radio, and sub-mm. This further suggests that the extended hard X-ray emission is associated with the vigorous nuclear star formation.

The soft (\hbox{0.5--1.5\,keV}) X-ray emission is shown in the bottom left panel of  Figure~\ref{obs3554}. It traces the inner spiral, emphasizing the northern spiral arm in contrast to the adjacent dust lane. The radio hotspots D, E, and G as well as their $^{12}$CO molecular counterparts fall within this dust lane and are absorbed in the soft X-rays. The southern spiral arm is much fainter and not visible in the X-ray band, because the NW part of the spiral is nearer to us, while the SE arm is seen absorbed through the galactic plane (see also Figure~\ref{HSTChandra}). The soft and hard X-ray emission peaks coincide in Figure~\ref{obs3554}. \citet{R05,R07} showed that the hard X-ray emission comes from the central AGN, which is always obscured by a neutral column of at least $\sim2\times10^{23}$\,cm$^{-2}$. This means that no direct soft X-ray emission is expected from the AGN below 1.5\,keV, implying either indirect AGN emission (scattering) or an extra unresolved starburst component which we discuss below. Considering the observed PSF profile shown in Figure~\ref{rprof}, the contributions from the wings of the PSF cannot dominate the observed soft emission from the center.

\subsection{Chandra spectra of the extended emission}

\begin{table*}
\begin{minipage}{17cm}
\caption{NGC\,1365 0.3--1.5\,keV emission: Fluxes and Spectroscopic Fits. All model fits include a cold absorber equal to the Galactic value of $1.4\times10^{20}$\,cm$^{-2}$; the second fit to ObsID\,3554 includes intrinsic absorption of $(7\pm2)\times10^{20}$\,cm$^{-2}$. The fluxes are quoted in units of $10^{-13}$\,erg\,s$^{-1}$\,cm$^{-2}$ and all \emph{XMM-Newton} fluxes are scaled by 20\% to account for the AGN contribution. All errors are 90\% confidence limits; fit parameters given in parenthesis were held fixed during the fit. Approximate values are given for parameters which were unconstrained by the fit.}
\begin{tabular}{@{}cccccccccc@{}}
\hline
ObsID & Counts & model & $F\si{0.3--1.5\,keV}$ &$\Gamma$ &$kT_1$ [keV] &$Z_1/Z_{\odot}$ &$kT_2$ [keV] & $Z_2/Z_{\odot}$ & $\chi^2$/DoF\\
\hline
3554              & 1591 & \emph{apec$+$zpow} &$5.3_{-1.1}^{+0.6}$  & (1) & $0.65\pm0.03$ & $0.17\pm0.04$ & ... & ...  & 48/54   \\
6871              & 2006 & \emph{apec$+$zpow} &$5.4_{-1.1}^{+0.5}$  & (1) & $0.64\pm0.02$ & $0.16\pm0.03$ & ... & ... & 43/46  \\
6872              & 2276 & \emph{apec$+$zpow} &$5.5_{-1.4}^{+0.7}$  & (1) & $0.64\pm0.02$ & $0.21\pm0.02$ & ... & ... & 50/56   \\
6873              & 2182 & \emph{apec$+$zpow} &$5.2_{-1.4}^{+0.5}$  & (1) & $0.65\pm0.03$ & $0.16\pm0.04$ & ... & ...  & 77/55   \\
6868              & 2221 & \emph{apec$+$zpow} &$5.4_{-1.0}^{+0.6}$  & (1) & $0.71\pm0.03$ & $0.15\pm0.03$ & ... & ...  & 69/59   \\
6869              & 2366 & \emph{apec$+$zpow} &$5.4_{-1.3}^{+0.5}$  & (1) & $0.63\pm0.02$ & $0.20\pm0.04$ & ... & ... & 83/55   \\
6870              & 2193 & \emph{apec$+$zpow} &$5.5_{-0.9}^{+0.5}$  & (1) & $0.62\pm0.02$ & $0.15\pm0.03$ & ... & ... & 66/53  \\
0151370101 & 9167 & \emph{apec$+$zpow} & $6.5_{-0.3}^{+0.3}$& (1) &  $0.66\pm0.01$  & $0.10\pm0.01$  & ... & ... & 348/318 \\ 
0151370201 & 2080 & \emph{apec$+$zpow} &$6.6_{-0.9}^{+0.8}$ & (1) & $0.70\pm0.03$ &  $0.09\pm0.01$  & ... & ... & 96/87 \\      
0151370701 & 4575 & \emph{apec$+$zpow} &$6.8_{-0.3}^{+0.4}$ & (1)  & $0.71\pm0.01$ & $0.10\pm0.01$ & ... & ... & 191/175 \\ 
0205590301 & 35873 & \emph{apec$+$zpow} &$6.8_{-0.1}^{+0.2}$ & (1) & $0.64\pm0.01$ & $0.096\pm0.004$ & ...  & ... & 702/626 \\  
0205590401 & 22517 & \emph{apec$+$zpow} &$6.8_{-0.4}^{+0.3}$ &  $1.0\pm0.5$ & $0.64\pm0.01$ & $0.10\pm0.01$  & ... & ... & 542/468 \\   
\hline
3554              & 1591 &  \emph{zwabs*apec} &$7.7_{-1.3}^{+0.6}$  & ... & $0.66\pm0.03$ & $0.07\pm0.01$ &  ... & ... &   52/54   \\
6871              & 2006 &  \emph{apec$+$apec} &$5.5_{-0.8}^{+0.4}$  & ... & $1.2\pm0.2$ & $0.06\pm0.03$ &   $0.61\pm0.04$ & (1) &   38/45   \\
6872              & 2276 &  \emph{apec$+$apec} &$5.4_{-0.4}^{+0.4}$  & ... & $1.0\pm0.1$ & (0.1) &  $0.55\pm0.04$ & (1) &  60/56   \\
6873              & 2182 &  \emph{apec$+$apec} &$5.4_{-0.7}^{+0.4}$  & ... & $0.99\pm0.07$ & $0.10\pm0.02$ & $0.54\pm0.07$ & (1) &    63/54   \\
6868              & 2221 &  \emph{apec$+$apec} &$5.4_{-0.6}^{+0.6}$  & ... & $1.7\pm0.4$ & $0.70\pm0.03$ & $0.70\pm0.03$ & (1)  &   68/58   \\
6869              & 2366 &  \emph{apec$+$apec} &$5.4_{-0.6}^{+0.5}$  & ... & $1.2\pm0.2$ & $0.10\pm0.03$ &  $0.58\pm0.04$ & (1) &    73/54   \\
6870              & 2193 &  \emph{apec$+$apec} &$5.4_{-0.7}^{+0.7}$  & ... & $1.2\pm0.2$ & $\la0.1$ &  $0.62\pm0.03$ & (1) &    68/52   \\
0151370101 & 9167 &  \emph{apec$+$apec} &$6.6_{-0.2}^{+0.4}$  & ... &  $1.5\pm0.1$  & $\la0.02$  &  $0.66\pm0.01$ & (1) & 350/317 \\ 
0151370201 & 2080 &  \emph{apec$+$apec} &$6.7_{-0.7}^{+0.6}$  &  ... & $1.3\pm0.2$ &  $\la0.02$  &   $0.66\pm0.03$ & (1) & 99/86 \\  
0151370701 & 4575 &  \emph{apec$+$apec} &$6.7_{-0.5}^{+0.4}$ & ... & $1.1\pm0.1$ & $0.02\pm0.01$ &   $0.66\pm0.02$ & (1) &190/174 \\ 
0205590301 & 35873 &  \emph{apec$+$apec} &$7.3_{-0.1}^{+0.2}$  &  ... & $1.7\pm0.1$ & $\la0.02$ &   $0.64\pm0.01$ & (1) &724/625 \\  
0205590401 & 22517 &  \emph{apec$+$apec} &$6.9_{-0.2}^{+0.1}$  &  ... & $1.15\pm0.05$ & $0.020\pm0.005$  &  $0.62\pm0.01$ & (1) & 547/468 \\   
\hline
\end{tabular}
\end{minipage}
\label{softTab}
\end{table*}

\emph{Chandra}'s spatial resolution allows us to model the extended emission of NGC\,1365 by excluding the AGN point source. As can be seen on Figure~\ref{rprof}, which shows the radial profile of NGC\,1365 during Obs. ID\,3554, the AGN point source dominates the emission from the inner $\sim2''$, which we excluded from the following analysis.  Figure~\ref{AllSpec} shows the \emph{Chandra} 2002 spectra extracted from the inner $2''$, a $40''$ circular annulus excluding the inner $2''$ (the extended emission), and the total emission within $40''$. The emission above $\sim3$\,keV is clearly dominated by the active nucleus, that below $\sim1.5$\,keV -- by the extended emission, as already noted by \citet{R05}. The extended emission peaks around 0.7--0.8\,keV, which suggest a thermal-plasma emission origin.  

After experimenting with a few different thermal-plasma models (XSPEC models \emph{mekal, cmekal, apec, raymond}, etc.), we find that \emph{apec} models the thermal emission best, but the differences are not significant.\footnote{All models include absorption equal to the Galactic value of $1.4\times10^{20}$\,cm$^{-2}$.} In the full \hbox{0.3--10\,keV} \emph{Chandra} band, in addition to the thermal-plasma component,  there is an extended power-law emission tail above 2\,keV. Using the 2--10\,keV extended-emission spectrum, we determined a best-fit photon index of $\Gamma=1.1\pm0.5$ in ObsID\,3554, which is substantially different from that of the AGN in the \hbox{2--10\,keV} band, whose simple power law slope is $\Gamma\approx-0.8$; this suggests that the extra component is not dominated by emission from the AGN.\footnote{In our original 2002 observation, the extended emission counts in the 2--10\,keV (5--10\,keV) band are equal to about 28\% (56\%) of the total AGN counts in the same band, while the expected contamination from the wings of the PSF should be about 10\%. In the subsequent \emph{Chandra} observations, when the AGN was more luminous, the contamination is stronger, especially around and above 5\,keV, which is also confirmed by comparing the hard-band spectral slope of the extended emission with that of the AGN.} The addition of the power-law component significantly imroves the \hbox{0.3--10\,keV} fits, with F-test probability of $10^{-8}$. The best fit \hbox{0.3--10\,keV} spectral model including the thermal plasma and power-law emission components for \emph{Chandra} ObsID\,3554 is shown in Figure~\ref{ChandraFullSpec}.  The \hbox{0.3--10\,keV} luminosities of each of the two components are: $3.0^{+1.1}_{-0.2}\times10^{40}$\,erg\,s$^{-1}$ for the thermal component and $1.8^{+0.9}_{-1.1}\times10^{40}$\,erg\,s$^{-1}$ for the power-law component. As noted above, the thermal plasma dominates the soft-band extended emission; in the \hbox{2--10\,keV} band the power law model contributes $1.4^{+0.4}_{-1.3}\times10^{40}$\,erg\,s$^{-1}$ to the extended emission, while the thermal plasma only $\sim1.3\times10^{39}$\,erg\,s$^{-1}$. 

\begin{figure}
\includegraphics[width=8cm]{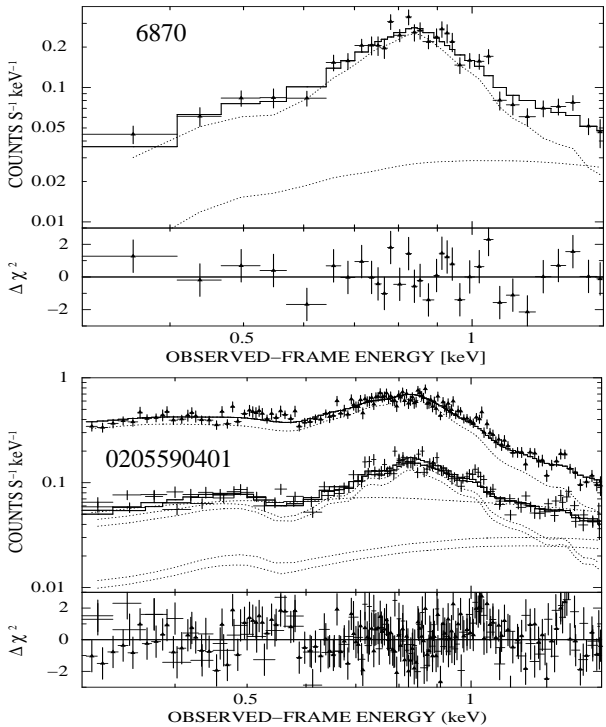}
\caption{Best fit 0.3-1.5\,keV spectra for an example \emph{Chandra} (Obs.\,ID~6870, top) and XMM-Newton (Obs.\,ID~0205590401, bottom) observation. The model fits include a thermal plasma (\emph{apec} in XSPEC) with a temperature of $\sim$0.62\,keV, and a power-law tail with a photon index $\Gamma=1$. The model fit components are shown separately with dotted lines below the fit; the fit residuals are shown in the lower panel.
\label{XraySpec}}
\end{figure}

For some of the remaining \emph{Chandra} observations, e.g. ObsID\,6868, the \hbox{2--10\,keV} emission seems to be contaminated by the AGN even beyond $2''$ from the center, as evidenced by the 2--10\,keV power law photon index of $\Gamma\approx-0.5$, the large ratio between AGN and extended-emission counts around 5\,keV, and the residuals between 6 and 7\,keV (which are consistent with an Fe-K$\alpha$ origin). Consequently, we restrict our analysis to the 0.3--1.5\,keV band, which, as shown above both spatially and spectroscopically, is dominated by the extended emission, even for the case of the much lower spatial resolution \emph{XMM-Newton} data. Table~\ref{softTab} lists the total counts and thermal plasma model fit parameters obtained in the soft band for all seven \emph{Chandra} observations.  The single temperature thermal plasma model provides a poor fit to the \hbox{0.3--1.5\,keV} data in all cases, and we consider the addition of a power law or a two-temperature plasma model instead. Both models provide significant improvements of the spectral fits (F-test probabilities of $<0.1$\%) and in most cases the current data does not show a clear preference for either of the two models. For the thermal plasma plus power-law model results shown in the top part of Table~\ref{softTab}, we fix the power law slope to $\Gamma=1$, the value found in the 2--10\,keV \emph{Chandra} fits unaffected by the AGN. This slope is confirmed in the soft band fit to the longest \emph{XMM-Newton} observation, as discussed below. An example \emph{Chandra} 0.3--1.5\,keV spectral fit from the upper part of Table~\ref{softTab} (Obs.\,ID~6870) is shown in Figure~\ref{XraySpec}.

Overall the \emph{Chandra} spectra of NGC\,1365 suggest a thermal plasma extended emission origin for the soft X-ray emission, with additional contribution from a second temperature thermal plasma or a $\Gamma=1$ power law. The fitted abundances in Table~\ref{softTab} are sub-solar, inconsistent with our general ideas about the recycling of materials around AGNs and the abundance measurements and gradients of NGC\,1365 (e.g., Galliano et al. 2005 report central abundance of 3 times solar). Sub-solar abundances are commonly found when applying simple thermal plasma models \citep[e.g.,][]{BF98}, and are considered a sign of the model simplifications (e.g. thermal equilibrium vs. shocks, etc.). 

From Figure~\ref{AllSpec}, the spectrum of the inner $2''$ emission has a low flux soft X-ray component extending below $2.5$\,keV. As noted in \S~\ref{spatial}, this unresolved core emission in the soft band is unlikely to originate directly from the active nucleus, which is absorbed by a large neutral column. The 0.3--1.5\,keV core (within $2''$) emission can be represented by a spectral model including thermal plasma ($kT=0.8\pm0.1$\,keV and abundances fixed to 0.17 of solar) and a power law absorbed by a Galactic neutral absorber consistent with the model fit of the extended emission. The luminosity of this core component is low -- $L\si{0.3--1.5\,keV}=4.4\pm1.1\times10^{39}$\,ergs\,s$^{-1}$, $\approx15\pm5\%$ of the 0.3--1.5\,keV band luminosity of the extended emission, too large to be attributed to the PSF wings of the extended emission surrounding the center.  Based on the similarity of the spectral shape of this core component to that of the extended star-formation associated emission, we speculate that this soft X-ray core component is dominated by unresolved recent star formation in the immediate vicinity of the active nucleus. 

\subsection{XMM-Newton spectra of the extended emission}
\label{XMMspec}

\begin{figure*}
\includegraphics[width=17cm]{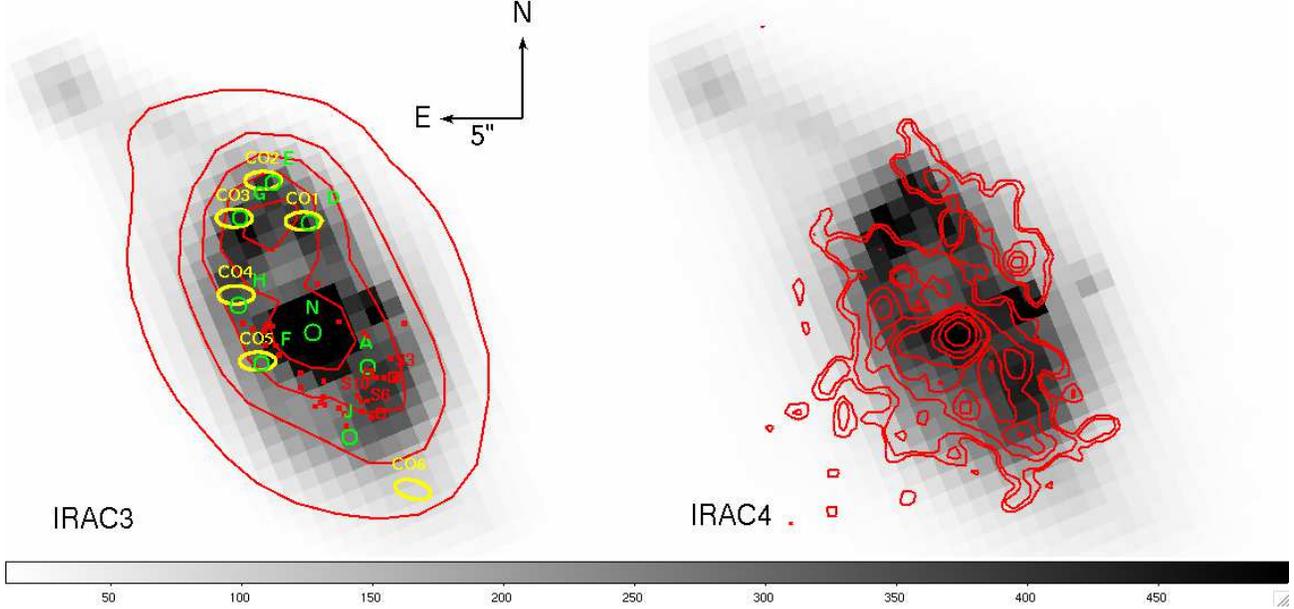}
\caption{\emph{Spitzer} IRAC3 (5.8\,$\mu$m, \emph{left}) and IRAC4  (8.0\,$\mu$m, \emph{right}) images of the nuclear starburst ring. The central nuclear emission dominates at and below $\sim$6\,$\mu$m; the circumnuclear starburst ring  emission is comparable at 8\,$\mu$m, and dominates the IR emission at longer wavelengths, for example, the 24\,$\mu$m MIPS emission shown in the left panel as contours. The right panel displays also the 0.3--10\,keV \emph{Chandra} contours (from  Obs. ID\,3554). The super-star clusters, radio hotspots, and $^{12}$CO molecular hotspots of the nuclear starburst ring are shown in the right panel with symbols following those of Figure~\ref{HSTChandra}.
\label{irac}}
\end{figure*}

Due to the fact that the soft emission is dominated by the extended component of NGC\,1365 and the heavily obscured AGN's contribution is expected to be negligible below 1.5\,keV, we can use the 0.3--1.5\,keV \emph{XMM-Newton} observations of NGC\,1365 to constrain the properties of the extended emission.  The unresolved soft X-ray emission component found within $2''$ of the active nucleus in the \emph{Chandra} observation above, has identical \hbox{0.3--1.5\,keV} spectrum and will only increase the total flux by about 15\%. In each case we extract the 0.3--1.5\,keV \emph{XMM-Newton} spectrum from a $40''$ region centered on NGC\,1365 and bin it to $>$20 counts per bin in order to use $\chi^2$ minimization fitting in XSPEC.  The corresponding \emph{XMM-Newton} spectral fits are given in Table~\ref{softTab}, together with the \emph{Chandra} fits for comparison. The quoted fluxes are \emph{XMM-Newton} $pn$ fluxes. Note that they are about 20\% brighter than the corresponding \emph{Chandra} fluxes, in agreement with the expected soft-band contribution of the unresolved core emission. The 0.3--1.5\,keV \emph{XMM-Newton} spectra are well represented by the thermal plasma plus power-law model found in the \emph{Chandra} fits, with the exeption of the two longest observations, ObsID\,0205590301 and ObsID\,0205590401, where this fit is inadequate ($\chi^2/DoF$=702/625 and $\chi^2/DoF$=542/468, respectively). Adopting varying abundances for some elements (e.g. Fe, Mg, Si, He) provides acceptable fits in both cases ($\chi^2/DoF$=663/621 and $\chi^2/DoF$=514/464), but considering the simplicity of the one-temperature collisionally ionized diffuse gas models, the caveats associated with the \emph{apec} and \emph{mekal} models in XSPEC, and the inconsistencies and poor constraints on the fitted abundance values of these model fits, we refrain from interpreting these abundance variations here. \citet{Wang09} suggest that allowing element abundances to vary independently for different spatial regions can resolve the problem of the small fitted overall abundance.

We checked the longest \emph{XMM-Newton} observation for varibility in the soft band (0.3--1.5\,keV) on timescales of hours, but found no significant variation.

\begin{figure}
\includegraphics[width=8cm]{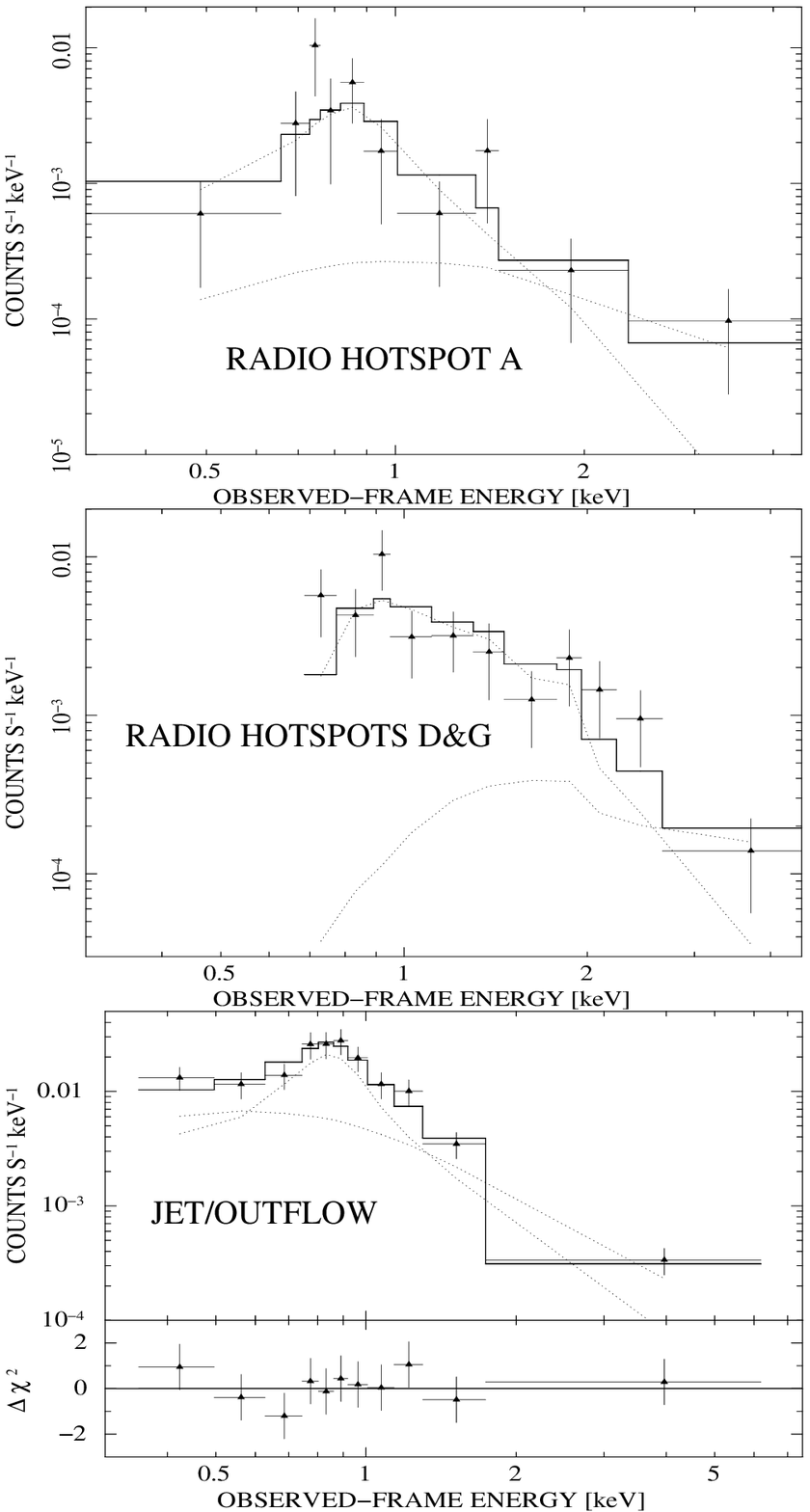}
\caption{X-ray spectra of select multi-wavelength counterparts for \emph{Chandra} Obs.\,ID~3554: the radio-hotspot A (SSC10; \emph{top panel}), radio hotspots D \& G (\emph{middle panel}), and radio hotspot F (jet/high-ionization outflow; \emph{bottom panel}).
\label{PartSpec}}
\end{figure}

\section{Discussion and Conclusions}

\subsection{Multiwalength View of the Nuclear Spiral}

The spatially-resolved \hbox{0.5--10\,keV} emission of NGC\,1365 traces well the nuclear spiral component, suggesting that the extended X-ray emission is associated with the circumnuclear starburst. This is confirmed by the spatially resolved \emph{Chandra} spectroscopy, which, together with the unresolved \emph{XMM-Newton} soft-band spectra, are dominated by  thermal plasma emission with temperature $\sim$0.7\,keV. The ultrasoft X-ray emission ($<$0.5\,keV), on the other hand, is confined to the outflow cone seen by \citet{HL96}, \citet{L99}, and \citet{Veilleux} in the kinematic signature of the high excitation optical emission lines. Soft-X-ray emission associated with a NLR outflow has been seen before, notably in NGC\,4151 and NGC\,1068, where the outflows are photoionized and photoexcited by the AGN \citep[e.g.,][]{Ogle,Ogle03,Schurch}. In the case of NGC\,1365, \citet{Veilleux} show that the optical narrow-line ratios associated with the outflow cone are consistent with AGN photoionization outside the nuclear spiral region, but have the photoionization characteristics of a starburst inside the inner 1--2\,kpc. In fact, as we argue below, the NGC\,1365 starburst  is powerful enough to drive a strong outflow like the one observed.

The molecular starburst  ring of NGC\,1365 is defined in the radio band, by radio 3\,cm and 6\,cm emission \citep{FN98,St99} and well traced in the MIR by \emph{Spitzer}, as well as by the string of super-star clusters in the optical \citep[SSCs;][]{K97}. There is evidence for molecular gas in the center which is dense ($\ga10^4$\,cm$^{-3}$) and cold \citep[kinetic temperature of 40\,K for NGC\,1365;][]{Ott05}, traced by NH$_3$, which correlates well with other star-formation indicators.  The CO emission maps of \citet{S07} confirm the presence of less dense molecular gas, with the $^{12}$CO and $^{13}$CO molecular maps tracing the circumnuclear ring structure dotted by emission hotspots. 

NGC\,1365 was recently imaged with \emph{Spitzer} with both IRAC and MIPS \citep{spitzer}. Figure~\ref{irac} shows the 5.8\,$\mu$m and 8\,$\mu$m IRAC images together with the 24\,$\mu$m MIPS (left panel) and the 0.3--10\,keV X-ray (right) contours, as well as the circumnuclear ring SSCs, radio, and  molecular CO hotspots. The circumnuclear star forming ring is traceable in the MIR in both panels,  gaining prominence with respect to the nuclear component at 8\,$\mu$m. At shorter MIR wavelengths the nuclear point source dominates the emission, at wavelengths increasingly longer than 8\,$\mu$m the starburst dominates. The 24\,$\mu$m MIPS emission contours shown in Figure~\ref{irac} peak in the NW-running dust lane, where the X-ray emission is mostly absorbed in the \emph{Chandra} band, but the molecular $^{12}$CO  emission (CO1, CO2, and CO3 in Figure~\ref{irac}) and the radio emission (hotspots D, E, and G) are strong. We address the to the nature of the emission in this strong MIR emission region in the next section.

The (spatially unresolved) 5--38\,$\mu$m spectrum of NGC\,1365 is a fairly shallow power law with no evidence of silicate absorption and weak 6.2, 7.7, 8.6, and 11.3\,$\mu$m 
polycyclic aromatic hydrocarbon (PAH) features. The starburst dominates, but a prominent AGN component must also be present, and overall the nucleus lies (fully on the starburst side but) close to the line separating starbursts and AGN on the MIR-slope diagnostic of \citet{spitzer}. Using our assumed distance of 21\,Mpc and the  \emph{Spitzer}  the 15\,$\mu$m and 30\,$\mu$m \emph{Spitzer} fluxes \citep{spitzer}, the MIR SFR estimate is $\rmn{SFR}_{Spitzer}\approx9$\,M$_{\odot}$\,yr$^{-1}$. This is about 25\% smaller than the FIR-estimated global NGC\,1365 SFR, which is $\rmn{SFR}\si{FIR}\approx12$\,M$_{\odot}$\,yr$^{-1}$, if we assume the Lonsdale et al.~1985 $L\si{FIR}=6.8\times10^{10}L_{{\odot}}$ measurement and eqn.~3 of \citet{K98}. 

The soft-band X-ray luminosity can also be used to estimate of the star-formation rate. With the help of eqn.~14 of \citet{XSFR}, $\rmn{SFR}=2.2\times10^{-40} L\si{0.5--2\,keV}$\,M$_{\odot}$\,yr$^{-1}$, and the observed \hbox{0.5--2\,keV} luminosity of the extended emission, $L\si{0.5--2\,keV}=3-3.6\times10^{40}$\,ergs\,s$^{-1}$ (where the higher value includes the contribution of an unresolved starburst component within 2$''$ from the AGN), we estimate a soft-X-ray  of $\rmn{SFR}\si{0.5--2\,keV}=7-8$\,M$_{\odot}$\,yr$^{-1}$. This is in good agreement (10--20\% lower) than the \emph{Spitzer} nuclear SFR estimate, which covers a slightly larger area, and about 30--40\% lower than the global FIR-estimated SFR. We conclude that most of the strong star formation (60--70\% or  7--8\,M$_{\odot}$\,yr$^{-1}$) is constrained to the nuclear starburst ring in the inner region of NGC\,1365. 

\subsection{Radio hotspots, Super-Star Clusters and Molecular clouds}

Radio hotspots A, D, G, and H have flat, non-thermal spectra between 6 and 20\,cm \citep{Sand95}. \citet{Sand95} suggest that radio hotspots A, D, and G (A and G are unresolved at 2\,cm, while D is only marginally detected) are related to the star formation and supernova activity. Radio hotspots D and G (which coincide with CO hotspots CO1 and CO3) fall within the prominent dark NW dust lane, and are associated with two obscured star-forming regions detected in the infrared by \citet{M98} and visible in the IRAC image of NGC\,1365 shown in \citet{spitzer} and our Figure~\ref{irac}. \citet{G05} and \citet{S07} suggest that the MIR peaks associated with radio hotspots D, E, and G \citep[M4, M5, and M6 in][]{G05,G08} correspond to SSCs. Using VLT images and spectra \citet{G08} revised the mass and age estimates of earlier works to show that the superstar clusters in NGC\,1365 are the most massive clusters found so far, with masses of at least $10^7$\,M$_{\odot}$ and ages of $\sim7$\,Myr. These SSCs are still embedded in large amounts of gas and dust due to dust trapping, unlike typical lower-mass star clusters found in less dense environments which sweep away their gaseous/dusty envelopes much earlier.

Using the X-ray images we can obtain estimates of the X-ray radiation from the radio and $^{12}$CO hotspots and the HST SSCs. X-ray spectra extracted from $1''$-regions\footnote{For \emph{Chandra} on-axis observations, 80--90\% of the total energy falls in a circle with radius $\sim1''$.} around the radio positions of sources D and G have only a handful of counts (16--24), but are consistent with the \emph{apec+zpow} (or simple \emph{apec}) models with parameters fixed to those reported in Table~\ref{softTab} for each \emph{Chandra} observation, if we also include an intrinsic absorber with column density of $\sim0.5-1\times10^{22}$\,cm$^{-2}$ (see the middle panel of Fig.~\ref{PartSpec}). The unabsorbed luminosity estimates for the 0.3--10\,keV band correspond to $L\si{x}\sim1-3\times10^{39}$\,erg\,s$^{-1}$ for both radio hotspots taken together. These luminosities are in agreement with the theoretical predictions for the highest mass clusters ($10^6$\,M$_{\odot}$) studied by \citet{Silich}, which have $L\si{x}\sim1-10\times10^{39}$\,erg\,s$^{-1}$ (depending on the density of the surrounding interstellar medium, the gas metallicity, etc.), but these predictions do not take into account the effects of gas-trapping in heavier clusters, which is likely relevant for the SSCs behind radio hotspots D and G.
 
Radio hotspot A, which has no molecular $^{12}$CO equivalent, falls in an optically and X-ray unabsorbed region and is likely associated with a super-star cluster SSC10 identified by \citet{K97} \citep{St99}.  There are 15--39 counts within $1''$ of the optical (similar for the slightly offset radio) positions in the different \emph{Chandra} observations and \emph{apec+zpow} Cash statistic model fits with parameters fixed to those of the extended \emph{Chandra} emission fits from Table~\ref{softTab} and free normalization provide acceptable fits in all cases (see the top panel of Fig.~\ref{PartSpec} for an example). The inferred 0.3--10\,keV $1''$ flux is $L\si{x}\sim6-10\times10^{38}$\,erg\,s$^{-1}$. The B-band and radio luminosities of SSC10 are $L\si{B}=4.2\times10^{39}$\,erg\,s$^{-1}$ and $L\si{3\,cm}=5.2\times10^{36}$\,erg\,s$^{-1}$,  the corresponding ratios $L\si{x}/L\si{B}\sim0.1-0.2$ and $L\si{3\,cm}/L\si{x}\sim0.005-0.009$. 

Radio hotspot F (which coincides with CO5) is traditionally identified with a radio jet \citep{Sand95,St99}. We argued above, that the ultrasoft X-ray emission is associated with the outflow cone seen in higher excitation lines like [O III] and [Ne III] \citep{HL96,L99}. We extracted X-ray spectra from the region between radio hotspots N and F, which overlaps with the ultrasoft emission contours shown in Figure~\ref{obs3554}, for each of the \emph{Chandra} pointings, excluding the $2''$-radius around the center which is affected by the AGN. The spectra can be fit by \emph{apec+zpow} models, with the \emph{apec} model parameters fixed to these quoted in Table~\ref{softTab} as shown in the bottom panel of Fig.~\ref{PartSpec}. Unlike the case of the overall extended emission component however, the power-law component dominates the  X-ray spectrum of the outflow at both high and low energies, i.e. the power-law index is much steeper -- typically $\Gamma\sim2.5$ -- and the normalization of this component much higher.  The \emph{apec+zpow} fits (including absorption equal to the Galactic value in all cases) are acceptable in most cases, but the residuals show excess emission below 0.5\,keV, i.e. the ultrasoft emission seen in Figure~\ref{obs3554} is not fully accounted for by the steeper power-law. The lower limit to the energy (since we exclude the highest surface brightness inner region due to AGN contamination) in the 0.3--10\,keV band for this outflow is $L\si{x,jet}\ga4\times10^{39}$\,erg\,s$^{-1}$. 

\subsection{Starburst-Driven Superwind}

Starburst-driven superwinds (as well as  AGN driven outflows) can explain the extended, often biconical, X-ray emission in nearby galaxies \citep[e.g.,][]{Heckman96,KS98}. \citet{KS98} concluded that the extended soft X-ray emission observed in NGC\,1365 could be the result of a starburst driven wind. We repeat the \citet{KS98}  estimates here, using the new  data on the SFR and new supernova (SN) rate estimates, assuming a velocity of $\sim$100--200\,km\,s$^{-1}$ on $\sim$1--2\,kpc scales.

We use two different methods to obtain an estimate of the SN rate.  \citet{SNrate} tabulate the expected SN rates as a function of B-band luminosity and the galaxy type. The B-band luminosity of NGC\,1365 is $7.9\times10^{10}L_{\odot}$ \citep[see Table~2 of][]{Blum}, resulting in a SN rate of 0.09\,yr$^{-1}$. Using the FIR luminosity method, we obtain slightly higher SN rate in the range 0.14--0.16\,yr$^{-1}$ \citep{vonBuren,Mannucci}. Both the B-band and the FIR-luminosity SN rate estimates apply to the whole galaxy.  Scaling the global SN rate by assuming that 60--70\% of the new stars and SN are born in the inner spiral, we arrive at a central SN rate in the range 0.05--0.1\,yr$^{-1}$. For comparison the radio-band inferred SN rate reported by \citet{St99} is about 0.02--0.03\,yr$^{-1}$ over a smaller ($\sim$1\,kpc$^2$) area including the central $\sim$0.6\,kpc. 

Following \citet{KS98}, the mechanical power or a starburst driven wind in NGC\,1365 is $L\si{mech}\approx1-3\times10^{42}$\,erg\,s$^{-1}$, assuming each supernova gives $L\si{SN}=10^{51}$\,erg\,yr$^{-1}$. With a soft-band luminosity of $L\si{0.1--2.4\,keV}=4.4\times10^{40}$\,erg\,s$^{-1}$, the \citet{KS98} estimates  \citep[their eqns. 1 through 3, based on][]{Heckman96} still hold, if we assume a mean density of the swept-up interstellar medium in the range $n=0.1-1$\,cm$^{-3}$ for an outflow velocity of $\sim$100--200\,km\,s$^{-1}$ on $\sim$1--2\,kpc scales. Therefore the observed starburst could supply the observed thermal X-rays in a wind driven shell. 

The H$\alpha$ emission expected for a SFR of $\sim$7-8\,M$_{\odot}$\,yr$^{-1}$ is $\sim10^{42}$\,erg\,s$^{-1}$ \citep{K98} much larger than the H$\alpha$ luminosity observed by \citet{K97}: $8\times10^{40}$\,erg\,s$^{-1}$ (converted to a distance of 21\,Mpc) for the two most luminous H$\alpha$ hotspots located SW of the nucleus (in the region of the HST-detected SSCs shown in Figures~\ref{HSTChandra}, \ref{obs3554}, and \ref{irac}). The total extended  H$\alpha$ luminosity  likely exceeds this estimate by a factor of a few. Considering the large obscuration inferred towards the star-formation regions in the NE of the nucleus (for example, those associated with radio hotspots D, and G, which are optically faint, but infrared and radio bright, hidden behind $N_H\approx10^{22}$\,cm$^{-2}$ according to the X-ray emission), it is conceivable that as little as $\sim$10\% of the H$\alpha$ luminosity associated with the vigorous star formation is observed directly.

\subsection{AGN contribution and IR-Optical-X-ray relations}

\citet{KS98} considered the place of the NGC\,1365 AGN among the IR-Optical-X-ray relations found by \citet{Ward88} for the sample of \citet{P82}. They concluded the AGN was too faint in the X-rays relative to the luminosity expected based on the observed broad H$\alpha$-line emission and the total IR emission. We know now that the AGN luminosity in the 2--10\,keV band, based on the \citet{Iyomoto97} fit to the ASCA data, was underestimated, and that in the Compton thin state, the AGN absorption-corrected luminosity is about an order of magnitude higher, $L\si{AGN,2--10\,keV}=0.8-1.7\times10^{42}$\,erg\,s$^{-1}$ \citep{R05,R07}. The value expected for the hard-band X-ray  luminosity based on the broad-line H$\alpha$ luminosity observed by \citet{Schulz94}, is $L\si{AGN,2--10\,keV}\approx0.6\times10^{42}$\,erg\,s$^{-1}$ (corrected to a distance of 21\,Mpc), comparable to the lower range of the observed hard-band AGN emission. 

Using the IR continuum fluxes at 15\,$\mu$m and 30\,$\mu$m given by \citet[][see their Table~3]{spitzer}, we estimate $L\si{25--60\,$\mu$m}=2.3\times10^{44}$\,erg\,s$^{-1}$ \citep[as in][including setting $H_0=50$\,km\,s$^{-1}$\,Mpc$^{-1}$ for the comparison]{Ward88}. The expected hard-band X-ray luminosities is then a factor of 10--20 higher than the \citet{R05b} observations. We believe this discrepancy is primarily due to the fact that the starburst dominates the IR emission beyond 6--15\,$\mu$m, while the \citet{Ward88} relations refer to the hard-band X-ray and IR emission of the AGN. According to \citet{Ward88}, the IR emission from the AGN component is NGC\,1365 should be $L\si{25--60\,$\mu$m}\la2\times10^{43}$\,erg\,s$^{-1}$. If the AGN dominates the emission at and below 6\,$\mu$m, the 6\,$\mu$m flux density should be a better predictor of the AGN hard-band flux. This is indeed the case. The average ratio of the 6\,$\mu$m flux density to the 2--10\,keV flux for type 2 AGNs was found to be $\log[F\si{2--10\,keV}/\nu F\si{$\nu$}(6\textrm{\,$\mu$m})]=-0.61$ by \citet{L04}, in agreement with the values observed for NGC\,1365: $-0.9\leq\log[F\si{2--10\,keV}/\nu F\si{$\nu$}(6\textrm{\,$\mu$m})]\leq-0.6$.

\section{Summary}

Using \emph{Chandra}'s excellent spatial resolution and \emph{XMM-Newton}'s higher photon gathering capabilities we study the extended nuclear emission of the supergiant barred galaxy NGC\,1365. The combination of radio, infrared, optical and X-ray data reveals the complex structure of the nuclear region (within about a kiloparsec of the center) highlighting the competing effects of the AGN, the circumnuclear starburst ring, and the central outflow. 

The ultrasoft X-ray emission (below $\sim$0.5\,keV) is spatially coincident with the AGN-driven outflow cone traced by higher excitation optical emission lines \citep{HL96,L99,Veilleux}. The X-ray spectrum of the outflow region can be represented by a dominant steep ($\Gamma\approx$2-3) power-law component; the steepness of the power law is driven by the excess emission below $\sim$0.5\,keV.

Spectroscopic as well as spatial evidence suggests that the soft X-ray emission (0.5--1.5\,keV) of NGC\,1365 has thermal origin related to the $\sim$1\,kpc-radius circumnuclear starburst. Thermal plasma emission models can fit the \emph{Chandra} and \emph{XMM-Newton} spectra well, with $kT\sim$0.6--0.7\,keV; a second thermal component with  $kT\sim$1.3\,keV or a power-law  with $\Gamma\approx1$ is also necessary to fit the observed spectra. Considering the high star-formation rate in the nuclear region ($\sim$7--8\,M$_{\odot}$\,yr$^{-1}$) and the associated high SN rates, a starburst-driven superwind is likely present, and can account for the wide angle outflow observed in the higher excitation optical emission lines and the ultrasoft X-rays.

The hard X-ray emission beyond $\sim$2\,keV is dominated by the obscured AGN, which also contributes substantially to the infrared emission below $\sim$6--15\,$\mu$m. The circumnuclear ring is also traced by low-surface brightness hard X-ray emission, suggesting that an obscuring column of  $\approx10^{22}$\,cm$^{-2}$ is hiding the $\sim7$\,Myr old (but still shrouded in gas and dust) $\sim10^7$\,M$_{\odot}$ super-star clusters seen in the radio, molecular CO, and mid-infrared in the NE part of the nuclear star-burst ring from view in the optical and soft X-ray bands.

The central regions of NGC\,1365 are influenced by the strong bar and nuclear spiral, which channel sufficient material to the innermost regions and are likely responsible for the circumnuclear starburst as well as fueling the supermassive black hole. Overall, the extended emission (which also contains unresolved point sources) contributes about $5\times10^{40}$\,erg\,s$^{-1}$ in the \emph{Chandra} X-ray band,  similar to the total emission from X-ray point sources \citep{SK09} during the 2002 \emph{Chandra} observation. The AGN is an order of magnitude more powerful  ($1-2\times10^{42}$\,erg\,s$^{-1}$) than the combined extended plus point source emission ($\sim10^{41}$\,erg\,s$^{-1}$), but $\sim$70\% of this emission is obscured by gas in the immediate vicinity of the active nucleus, which is associated with the broad-line clouds.  The moderate-luminosity AGN is often outshined by the nuclear starburst (e.g., in the radio and 24\,$\mu$m infrared, as well as soft 0.5--1.5\,keV X-ray emission), and both the infrared and X-ray emission result from the complex interplay of the stellar and AGN emission modified by intrinsic absorption, as well as absorption related to orientation of the inner spiral with respect to our line of site.

\section*{Acknowledgements}

This project has been partly funded by the DFG Priority Programme 1177 "Galaxy Evolution''. We thank Peter Predehl for useful discussions and a critical reading of the manuscript.

This work is based on observations obtained with \emph{XMM-Newton}, an ESA science mission with instruments and contributions directly funded by ESA Member States and the US (NASA). In Germany, the  \emph{XMM-Newton} project is supported by the Bundesministerium f\"ur Wirtschaft und Technologie/Deutsches Zentrum f\"ur Luft- und Raumfahrt (BMWI/DLR, FKZ 50 OX 0001) and the Max-Planck Society. It is also based on an observation obtained with the \emph{Chandra} X-ray telescope (a NASA mission).

%%%%%%%%%%%%%%%%%%%%%
%%%%%%% FIGURES %%%%%%%%%
%%%%%%%%%%%%%%%%%%%%%

%%%%%  %%%%%%

%%%%%  %%%%%%

\appendix

\section{MS\,0331.3$-$3629}

\begin{table}
\begin{minipage}{81mm}
\caption{MS\,0331.3$-$3629: \emph{Chandra}/\emph{XMM-Newton} Observation Summary. (1) Observation ID; (2) Modified Julian Day; (3) Total counts; (4) Effective exposure time, after corrections for flaring and chip position.}
\begin{tabular}{@{}cccc@{}}
\hline
ObsID & MJD & Counts & T$\si{eff}$\\
(1) & (2) &(3) &(4)\\
& &0.3--10\,keV &$10^3$\,s\\
\hline
                                  &              & ACIS-I2  &  ACIS-I2\\
\hline
3554         		      & 52632 &  2758     &  14.6 \\
\hline
                                  &              & pn  MOS1 MOS2  & pn  MOS1 MOS2 \\
\hline
0151370101 & 52655 &  5888  2767 2700   &  12.7 17.4 17.4 \\ 
0151370201 & 52679 &    ...         712   675  &   ...   5.1   5.0  \\ 
0151370701 & 52864 &    ...         843   754  &   ...   8.1   8.2 \\ 
0205590301 & 53021 & 18088 6006  6631 &  48.7 57.3  57.4 \\
0205590401 & 53210 & 8700 4182  3713 &  28.6 46.4  47.1 \\
\hline
\end{tabular}
\label{tabChandraXMM}
\end{minipage}
\end{table}

\subsection{\emph{Chandra} and \emph{XMM-Newton} Imaging Spectroscopy of MS\,0331.3$-$3629}

\begin{table*}
\begin{minipage}{122mm}
\caption{MS\,0331.3$-$3629: 0.3--10\,keV Band Single Power-law Spectroscopic Fits. (1) Observation ID; (2) Hard-band rest-frame absorption-corrected ACIS-S or MOS/pn average fluxes in units of  $10^{-13}$\,erg\,s$^{-1}$\,cm$^{-2}$; (3) Soft-band rest-frame absorption-corrected ACIS-S or MOS/pn average fluxes in units of  $10^{-13}$\,erg\,s$^{-1}$\,cm$^{-2}$; (4) Power-law photon index; (5) Intrinsic abroption in units of $10^{20}$\,cm\,s$^{-1}$ (6) Model fit: ``PL'' indicates a power law fit [wabs*zpow in XSPEC], ``APL'' indicates a power law fit including intrinsic absorption [wabs*(zwabs*zpow) in XSPEC]. A cold absorber equal to the Galactic value, $N\si{H}=1.4\times10^{20}$\,cm$^{-2}$, is included in the fits in all cases; (7) Data used in the model fits; (8) $\chi^2$ per degree of freedom (DoF).}
\begin{tabular}{@{}cccccccc@{}}
\hline
ObsID &$F\si{2--10\,keV}$ &$F\si{0.5--2\,keV}$ &$\Gamma$ &$N\si{H,i}$ &Model & Data &$\chi^2$/DoF\\
(1) &(2) &(3) &(4) &(5) &(6) &(7) &(8)\\
\hline
3554              & $19.6_{-0.1}^{+0.1}$  & $20.5_{-0.2}^{+0.2}$  & 2.06$\pm$0.04 & ...                    &    PL & ACIS-S     & 123/111   \\
0151370101 & $20.8_{-0.5}^{+0.7}$ & $16.8_{-0.5}^{+0.4}$  & 1.95$\pm$0.02 & ...                    &    PL & pn+MOS & 490/462 \\
0151370201 & $15.3_{-1.6}^{+1.4}$ & $17.8_{-1.2}^{+1.4}$  & 2.00$\pm$0.05 & ...                    &    PL & MOS        & 60/63 \\ 
0151370701 & $16.4_{-1.3}^{+1.4}$ & $17.0_{-1.1}^{+1.4}$  & 2.08$\pm$0.04 & ...                    &    PL & MOS        & 78/77 \\
0205590301 & $14.9_{-0.6}^{+0.5}$ & $18.1_{-0.5}^{+0.5}$  & 1.97$\pm$0.01 & ...                    &    PL & pn+MOS & 967/953 \\  
0205590401 & $16.2_{-0.6}^{+0.5}$ & $16.6_{-0.5}^{+0.5}$  & 2.08$\pm$0.01 & ...                    &    PL & pn+MOS & 622/664 \\  
\hline 
0205590301 & $16.2_{-0.4}^{+0.4}$ & $17.6_{-0.2}^{+1.0}$  & 2.05$\pm$0.02 & 2.4$\pm$0.5 & APL & pn+MOS & 939/952 \\  
0205590401 & $17.4_{-0.4}^{+0.4}$ & $16.0_{-0.2}^{+1.0}$  & 2.16$\pm$0.02 & 2.0$\pm$0.1 & APL & pn+MOS & 612/663 \\  
\hline
\end{tabular}
\end{minipage}
\label{tabFullFits}
\end{table*}

We present a series of high-quality (up to $\sim2\times10^4$ counts) X-ray imaging spectra of the high-peaked BL Lac (HBL) MS\,0331.3$-$3629. MS\,0331.3$-$3629 (03:33:12.2, $-$36:19:48.0) is a $z=0.308$ blazar candidate \citep{Stocke91} which was serendipitously detected in six recent long exposures of NGC\,1365 
-- falling within one \emph{Chandra} and five \emph{XMM-Newton} fields. Table~\ref{tabChandraXMM} shows a summary of the recent \emph{Chandra} and \emph{XMM-Newton} observations of MS\,0331.3$-$3629. We used standard \emph{Chandra} and \emph{XMM-Newton} pipeline reductions to obtain X-ray spectra (grouped into bins of at least 20 counts per bin) of the blazar from an elliptical region with a semi-major axis of $90''$ for the pn and $60''$ for the MOS cameras (for an off axis angle $\sim 12.5'$). About half of the exposure time of ObsID\,0205590401 is lost to flares, while for ObsID\,0151370201 and ObsID\,0151370701 the BL Lac falls within the chip-gaps of the pn camera, and we use the MOS data only.

\begin{figure*}
\includegraphics[width=17cm]{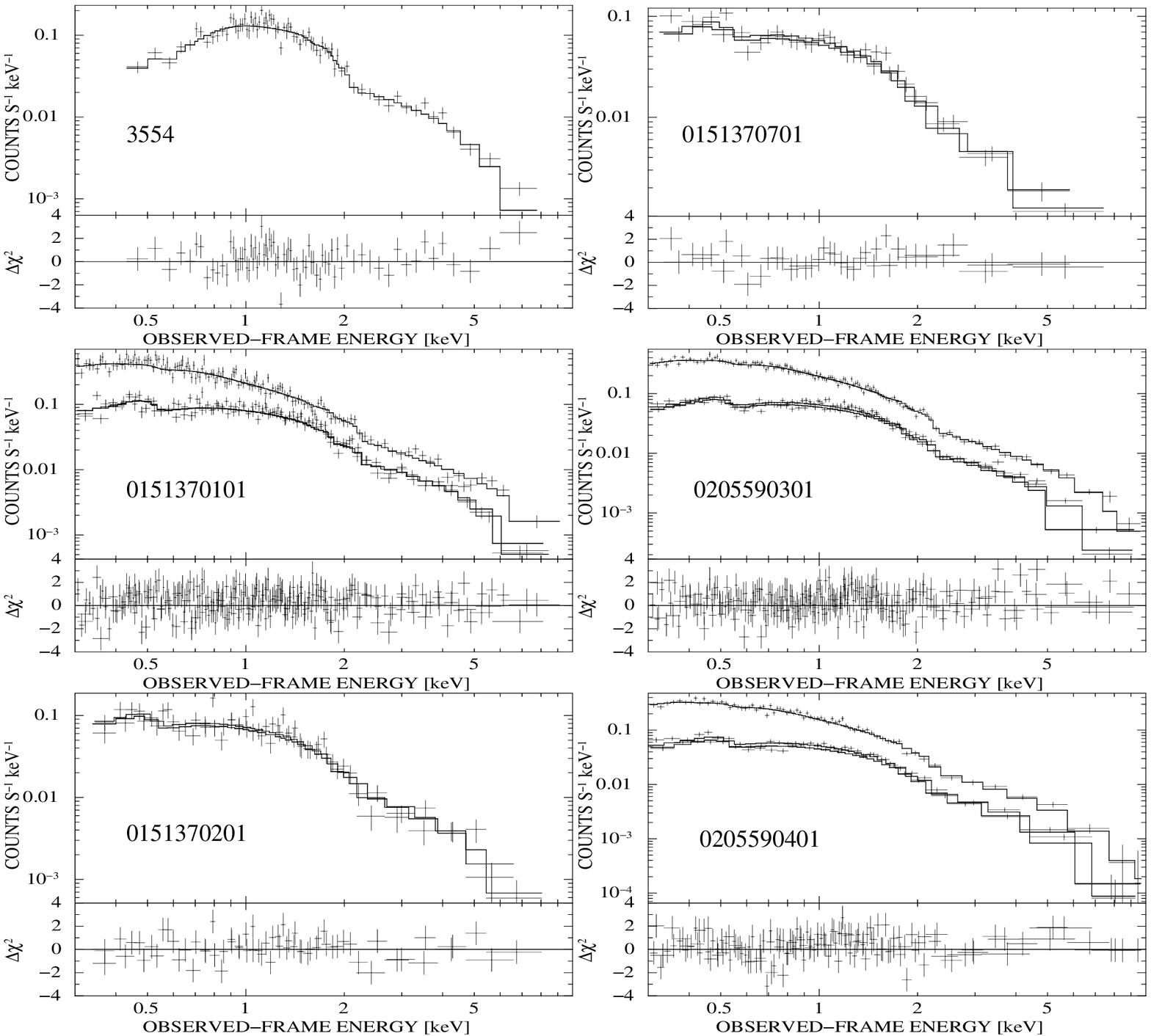}
\caption{The \hbox{0.3--10\,keV} spectral fits of MS\,0331.3$-$3629. In each case, the top panel shows the spectrum and best fit model from Table~\ref{tabFullFits} as well as the ObsID, the bottom panel -- the fit residuals. 
\label{BLXraySpec}}
\end{figure*}

The 0.3--10\,keV spectra can be fit by single power-law models modified by Galactic absorption. The fit parameters are given in the top portion of Table~\ref{tabFullFits}. 
The fits to the two highest quality  \emph{XMM-Newton} spectra can be improved by adding an intrinsic absorber $N\si{H,i}\approx 2 \times 10^{20}$\,cm$^{-2}$ (F-test probability of $1\times10^{-7}$ and $2\times10^{-3}$ for ObsID\,0205590301 and ObsID\,0205590401, respectively; last two rows of Table~\ref{tabFullFits}). The best spectral fits of all spectra (including intrinsic absorption in the case of the two longest  \emph{XMM-Newton} exposures) are shown in  Figure~\ref{BLXraySpec}. Broken power-law fits (\emph{bknpow} in XSPEC) are also consistent with the \emph{XMM-Newton} observations (an F-test shows no improvement to the model fit for the \emph{Chandra} observation) and the fit results are presented in Table~\ref{bknpow}. For the five \emph{XMM-Newton} fits, the break energy is around \hbox{$\left<E\si{break}\right>\approx1.1$\,keV}\footnote{$\left<\right>$ indicate weighted averages.}, the soft photon index \hbox{$\left<\Gamma_1\right>\approx1.9$}, and hard photon index \hbox{$\left<\Gamma_2\right>\approx2.1$}, although all values show large variations between the different observations. The current data cannot distinguish between the intrinsic absorption and broken power-law fit scenarios; the broken power-law fits have one extra parameter but do not provide statistically better fits in comparison to the intrinsically-absorbed fits, except for Obs\,ID~0205590401 (an F-test probability of $2\times10^{-3}$).

\subsection{X-ray Spectral Curvature of MS\,0331.3$-$3629}

\citet{curve} suggest that a large fraction of BL Lacs show spectral curvature in the \hbox{0.5--10.0\,keV} band, which can be detected using spectral fits in consecutive energy bands. This spectral curvature requires a change in the standard energy emission mechanism (which produces a simple power-law), and \citet{curve} propose to explain this curvature by an episodic (or time-variable) particle acceleration process. Using the highest signal-to-noise best \emph{XMM-Newton} spectrum of MS\,0331.3$-$3629 (Obs\,ID~0205590301), we looked for signs of curvature similar to those found by \citet{curve}. MS\,0331.3$-$3629 shows no clear signs of spectral curvature in the 0.5--10\,keV range. In addition, with the exception of Obs\,ID~0205590401, the broken power-law fit results, which are not preferable to a simple absorbed power law, suggest that varying the photon index with energy bandpass is not necessary to describe the X-ray spectra. Following \citet{curve}, this could mean that the radio to soft-X-ray emission of MS\,0331.3$-$3629 is consistent with synchrotron radiation produced by electrons with a simple power-law distribution.

\subsection{Previous X-ray observations of MS\,0331.3$-$3629}

MS\,0331.3$-$3629 was serendipitously detected in eight \emph{ROSAT} observations (see Table~\ref{tabRosat}). We performed source detection and photometry on the 4 PSPC (channels 52--201, corresponding to 0.5--2.0\,keV) and 3 HRI (channels 0--15, 0.1--2.4\,keV) event files using EXSAS. The resulting count rates and 0.5--2.0\,keV fluxes (assuming $\Gamma=2$) are given in Table~\ref{tabRosat}. The RASS detection quoted in Table~\ref{tabRosat} was obtained from the \emph{ROSAT} bright source catalog \citep{voges99}. \citet{Lamer96} used the longest PSPC exposure to measure a photon index in the 0.1--2.4\,keV band, $\Gamma\si{soft}=2.1\pm0.15$. Table~\ref{tabROSATspec} gives the soft-band (0.1--2.5\,keV in the case of \emph{ROSAT} PSPC) spectral fits obtained with XSPEC in a manner identical to the one used for the \emph{Chandra}/\emph{XMM-Newton} X-ray spectra. We used standard EXSAS processing to obtain the spectra, varying the size of the extraction region to include 90\% of the flux in each case. A simple power-law fit (indicated with ``PL'' in Table~\ref{tabROSATspec}) with absorption equal to the Galactic value and a photon index between 1.8 and 2.1 is sufficient to represent the soft-band \emph{ROSAT} PSPC spectra in three of the four cases. In the case of RP700921A01, the PL fit is only marginally acceptable ($\chi^2/DoF =53/40$, null hypothesis probability of 8\%). The inclusion of intrinsic absorption does not improve the model fits.
 
MS\,0331.3$-$3629 was also observed by \emph{BeppoSAX} (S/N=3.3) and is part of the HELLAS sample \citep{hellas}. \citet{emss} show the optical spectrum of MS\,0331.3$-$3629 as part of the \emph{Enstein} medium sensitivity survey BL Lac catalog. The \emph{Enstein} flux was $F\si{0.3-3.5\,keV}=5.22\times10^{-13}$\,erg\,s$^{-1}$\,cm$^{-2}$, consistent with the values listed in Table~\ref{tabROSATspec}.

\subsection{X-ray variability of MS\,0331.3$-$3629}

The MS\,0331.3$-$3629 X-ray flux is variable on timescales of weeks to years. The soft (\hbox{0.5--2.0\,keV}) band variability is about a factor of 2 over 7.5 years (rest-frame); the hard (\hbox{2--10\,keV}) band variations appear smaller, but the time interval probed is also shorter. Only the \emph{Chandra} observation shows short term count-rate variability of about 20\% on timescale of hours, but the evidence for this short term variability is not consistent: the Kolmogorov-Smirnov test confirms it with a 97\% confidence, but a $\chi^2$ test  gives a $\sim$60\% probability of constancy. The power-law photon index also varies between the observations on timescales of days to years.

\subsection{Multiwavelength Observations and the Nature of MS\,0331.3$-$3629}

The optical spectrum of MS\,0331.3$-$3629 shows no emission lines and has a Ca~II H\&K break of 30\% \citep{Stocke91}, consistent with a BL Lac classification at $z=0.308$.  An HST observation, taken on 1996 October 29,  revealed an elliptical host which outshines the active nucleus at 7020\AA\ \citep[the host apparent magnitude is $R=17.83$, compared to $R=19.03$ for the active nucleus,][]{HST}. Note that most prior work on the optical-to-X-ray and optical-to-radio flux ratios used the total galaxy magnitude in the optical, leading to incorrect estimates. The radio-to-X-ray index, defined as $\alpha\si{rx}=-\log[F\si{5GHz}/F\si{1keV}]/\log[\nu\si{5GHz}/\nu\si{1keV}]$, is $\alpha\si{rx}=0.55$. 

MS\,0331.3$-$3629 has been observed in the J, H, and K bands \citep[2\,MASS, likely dominated by the host galaxy,][] {2mass}, and the radio: the VLA \citep{Stocke91}, NVSS \citep[observed on 1993-10,][]{nvss}, SUMMS \citep[1998/12--2002/08 composite,][]{summs}. If we construct a spectral energy distribution (SED) based on these heterogeneous and non-concurrent observations, it peaks at $\log (\nu_p) \approx17.0$, and MS\,0331.3$-$3629 has the appearance of a high energy peaked blazar \citep{XrayBL}. Assuming that the radio to soft X-ray emission is dominated by synchrotron emission, a log-parabolic fit to the SED is possible \citep[see for e.g.,][]{Massaro1,Massaro2}.  Under the log-parabolic model, the bolometric luminosity of the synchrotron component can be estimated analytically \citep[e.g.,][]{Perri,Massaro2}, \hbox{$L\si{Synch} \approx 4\times10^{45}$\,erg\,s$^{-1}$}.

%% appendix tables

\begin{table*}
\begin{minipage}{112mm}
\caption{MS\,0331.3$-$3629: 0.3--10\,keV Band Broken Power-law Spectral Fits. (1) Observation ID; (2) Hard-band rest-frame absorption-corrected ACIS-S or MOS/pn average fluxes in units of  $10^{-13}$\,erg\,s$^{-1}$\,cm$^{-2}$; (3) Soft-band rest-frame absorption-corrected ACIS-S or MOS/pn average fluxes in units of  $10^{-13}$\,erg\,s$^{-1}$\,cm$^{-2}$; (4) First and (5) Second power-law photon indices; (6) Break energy, in keV;  (7) $\chi^2$ per degree of freedom (DoF).}
\begin{tabular}{@{}ccccccc@{}}
\hline
ObsID &$F\si{2--10\,keV}$ &$F\si{0.5--2\,keV}$ &$\Gamma_1$ &$\Gamma_2$  &$E\si{break}$ &$\chi^2$/DoF\\
(1) &(2) &(3) &(4) &(5) &(6) &(7)\\
\hline
3554              & $20.6_{-0.1}^{+0.2}$  & $15.4_{-1.0}^{+1.8}$ & 2.16$\pm$0.08 & 1.95$\pm$0.11 & 1.7$\pm$0.7 & 120/109   \\
0151370101 & $19.3_{-0.3}^{+0.4}$ & $16.0_{-1.0}^{+0.7}$ & 1.91$\pm$0.04 & 2.01$\pm$0.04 & 1.3$\pm$0.5  & 486/460 \\
0151370201 & $16.5_{-3.7}^{+3.2}$ & $15.4_{-1.0}^{+1.8}$ & 1.87$\pm$0.12 & 2.2$\pm$0.2 & 1.4$\pm$0.6  & 57/61 \\ 
0151370701 & $16.9_{-2.9}^{+6.3}$ & $15.5_{-2.4}^{+1.9}$ & 2.03$\pm$0.06 & 2.3$\pm$0.3 & 2.4$\pm$1.5  & 76/75 \\
0205590301 & $15.2_{-0.9}^{+0.9}$ & $17.1_{-0.7}^{+0.7}$ & 1.87$\pm$0.03 & 2.05$\pm$0.02 & 1.1$\pm$0.1  & 937/951 \\
0205590401 & $16.6_{-0.9}^{+0.9}$ & $15.2_{-0.7}^{+0.7}$ & 1.99$\pm$0.03 & 2.20$\pm$0.04 & 1.2$\pm$0.2  & 605/662 \\  
\hline
\end{tabular}
\label{bknpow}
\end{minipage}
\end{table*}

\begin{table*}
\begin{minipage}{112mm}
\caption{MS\,0331.3$-$3629: \emph{ROSAT} Photometry. (1) \emph{ROSAT} observation ID, where ``RP" refers to the PSPC, ``RH" to the HRI, and ``RS" to the \emph{ROSAT} all sky survey; (2) Observation date; (3) Effective exposure time; (4) Off-axis angle; (5) Count rate in the observed 0.5--2.0\,keV band (PSPC) or 0.1--2.4\,keV band (HRI); (6) Rest-frame 0.5--2.0\,keV absorption-corrected flux.}
\begin{tabular}{@{}cccccc@{}}
\hline
ObsID &Date &T$\si{eff}$ &OffAngle & CR & $F\si{0.5--2\,keV}$\\
(1) &(2) &(3) &(4) & (5) & (6)\\
 & yy-mm-dd & s & $'$& counts\,s$^{-1}$& $10^{-13}$\,erg\,s$^{-1}$\,cm$^{-2}$\\
\hline
RP800301N00 & 1992-08-22 &  840	& 43.1 & 0.087$\pm$0.011 & 10$\pm$1.3  \\ 
RP700921N00 & 1992-08-23 & 2323	& 12.4 & 0.090$\pm$0.006 & 11$\pm$0.7  \\ 
RP800301A01 & 1993-02-02 & 4121	& 43.1 & 0.048$\pm$0.004 & 5.7$\pm$0.5 \\ 
RP700921A01 & 1993-02-10 & 6908	& 12.4 & 0.062$\pm$0.003 & 7.5$\pm$0.4 \\ 
RH701297N00 & 1994-08-04 & 9182	& 12.4 & 0.037$\pm$0.002 & 7.9$\pm$0.5 \\ 
RH701297A01 & 1994-08-23 &  323	& 12.4 & 0.032$\pm$0.011 & 6.8$\pm$2.3 \\ 
RH701297A02 & 1995-07-04 & 9074	& 12.4 & 0.037$\pm$0.002 & 7.9$\pm$0.5 \\ 
RS932309N00 & 1990-07-30 &  172         & ...      & 0.090$\pm$0.028 & 5.2$\pm$1.7 \\ % rasbsc
\hline
\end{tabular}
\label{tabRosat}
\end{minipage}
\end{table*}

\begin{table*}
\begin{minipage}{120mm}
\caption{MS\,0331.3$-$3629: Soft-band Spectroscopic Fits. (1) Observation ID; (2) Soft-band power-law photon index; (3) Rest-frame 2\,keV monochromatic flux density; (4) Rest-frame absorption-corrected fluxes; (5) Bandpass used in the model fit; (6) Model fit: ``PL'' indicates a power law fit [wabs*zpow in XSPEC], ``APL'' indicates a power law fit including intrinsic absorption [wabs*(zwabs*zpow) in XSPEC] with $N\si{H,i}=2.0\pm0.5\times10^{20}$\,cm$^{-2}$. A cold absorber equal to the Galactic value, $N\si{H}=1.4\times10^{20}$\,cm$^{-2}$ is included in the fits in all cases; (7) $\chi^2$ per degree of freedom (DoF). }
\begin{tabular}{@{}ccccccc@{}}
\hline
ObsID &  $\Gamma\si{soft}$   &  $f\si{2\,keV}$ &  $F\si{0.5--2\,keV}$   & Data & Model  & $\chi^2$/DoF \\
(1) & (2) & (3) &(4) & (5) &(6) &(7)\\
& & $10^{-8}$\,Jy & $10^{-13}$\,erg\,s$^{-1}$\,cm$^{-2}$ & keV & & \\
\hline
RP800301N00 & 1.81$\pm$0.28  & 2.9$\pm$0.7   & $17_{-7}^{+8}$          & 0.1--2.4 &  PL  &  6/7 \\
RP700921N00 & 1.91$\pm$0.09  & 2.9$\pm$0.3   & $18_{-3}^{+2}$          & 0.1--2.4 &  PL  & 31/37 \\
RP800301A01 & 1.91$\pm$0.12  & 1.8$\pm$0.3   & $11_{-2}^{+3}$          & 0.1--2.4 &  PL  & 19/20\\
RP700921A01 & 2.10$\pm$0.05  & 1.9$\pm$0.2   & $14_{-2}^{+2}$          & 0.1--2.4 &  PL  & 53/40 \\
3554                  & 2.12$\pm$0.06  & 2.7$\pm$0.2 & $19.9_{- 2.0}^{+1.9 }$ &  0.2--2.5 & PL   & 97/87 \\
0151370101     & 1.92$\pm$0.02  & 2.7$\pm$0.1 & $15.9_{- 1.1}^{+1.2 }$ & 0.2--2.5  & PL   & 427/411\\
0151370201     & 1.87$\pm$0.06  & 2.5$\pm$0.2 & $15.1_{- 2.3}^{+2.3 }$ & 0.2--2.5  & PL   & 59/55\\
0151370701     & 2.05$\pm$0.06  & 2.4$\pm$0.2 & $16.6_{- 1.9}^{+2.1}$  & 0.2--2.5  & PL   & 72/71\\
0205590301     & 2.04$\pm$0.03  & 2.3$\pm$0.1 & $16.0_{- 1.1}^{+1.0}$  & 0.2--2.5  & APL & 802/784\\
0205590301     & 1.93$\pm$0.01  & 2.3$\pm$0.1 & $14.8_{- 0.6}^{+0.6}$  & 0.2--2.5  & PL    & 821/785\\
0205590401     & 2.06$\pm$0.02  & 2.3$\pm$0.1 & $16.3_{- 0.9}^{+0.8}$  & 0.2--2.5  & PL    & 517/573\\
\hline
\end{tabular}
\label{tabROSATspec}
\end{minipage}
\end{table*}

\label{lastpage}

\end{document}